%% file: 2019-adm-arxiv.tex
\documentclass{article}

\usepackage[margin=0.95in]{geometry}
\usepackage{amsmath, mathtools, mathrsfs}
\usepackage{graphicx}
\usepackage{import}
\usepackage{color}
\usepackage{hyperref}
\usepackage{authblk}

\newcommand{\dotted}{\protect\mbox{${\mathinner{\cdotp\cdotp\cdotp\cdotp\cdotp}}$}}
\newcommand{\full}{\protect\mbox{------}}

\begin{document}

\title{Filtered actuator disks: Theory and application to wind turbine models in large eddy simulation}
\author{Carl R. Shapiro, Dennice F. Gayme, and Charles Meneveau}
\affil{Department of Mechanical Engineering \\ Johns Hopkins University \\ Baltimore, Maryland, USA}
\date{}

\maketitle

\begin{abstract}
The actuator disk model (ADM) continues to be a popular wind turbine representation in large eddy simulations (LES) of large wind farms. Computational restrictions typically limit the number of grid points across the rotor of each actuator disk and require spatial filtering to smoothly distribute the applied force distribution on discrete grid points. At typical grid resolutions, simulations cannot capture all of the vorticity shed behind the disk and subsequently over-predict power by upwards of 10\%. To correct these modeling errors, we propose a vortex cylinder model to quantify the shed vorticity when a filtered force distribution is applied at the actuator disk. This model is then used to derive a correction factor for numerical simulations that collapses the power curve for simulations at various filter widths and grid resolutions onto the curve obtained using axial momentum theory. The proposed correction therefore facilitates accurate power measurements in LES without resorting to highly refined numerical grids.
\end{abstract}

\section{Introduction}
Blade-resolved simulations of wind turbines~\cite{Vijayakumar2016a}, which can directly calculate the lift, drag, loading, and deflection of the blades, are computationally costly and thus mostly limited to simulations of a single wind turbine. In wind farm simulations, turbines are instead represented by actuator models, such as the actuator line model (ALM)~\cite{Sorensen2002a, Martinez2018a} and the actuator disk model (ADM)~\cite{Burton2011a}. Detailed simulations of wake characteristics and loading are provided by ALM, which models the lift and drag along the blades as a distribution of forces~\cite{Sorensen2002a, Martinez2018a}. However, the coarse grid resolution needed when simulating very large wind farms, which are becoming more prevalent,~\cite{Allaerts2018a} or applying optimal-control methods~\cite{Goit2015a} are often too costly for ALM. As a result, ADM, which distributes the thrust force of the turbine across the swept area of the rotor, remains a popular wind turbine representation~\cite{Iungo2017a, Stevens2018a, Lignarolo2016a}.

\begin{figure}
\begin{center}
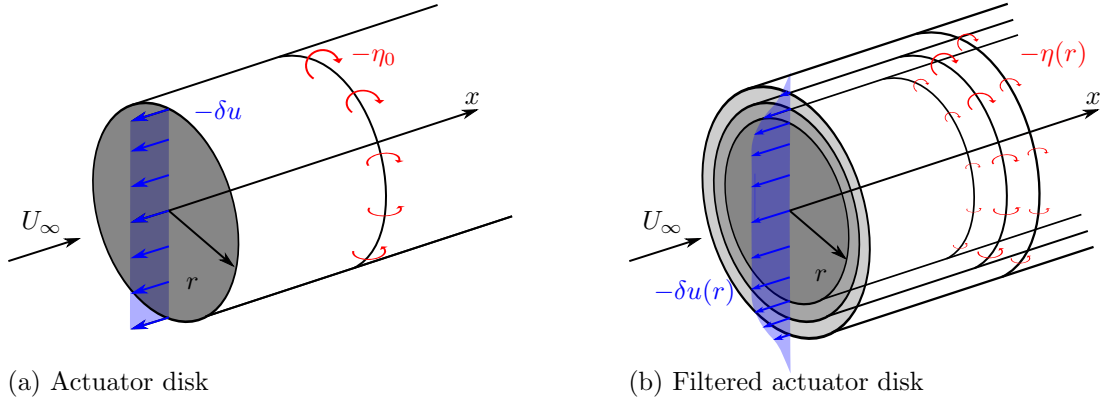
\caption{\label{fig:diagram} (a) Shed circulation $\eta_0$ per unit length of the vortex cylinder behind an actuator disk (shown in gray) and velocity induced at the disk $\delta u$. (b) Shed circulation distribution per unit area $\eta(r)$ of the concentric vortex cylinders shed by a filtered actuator disk (shown in gray). The velocity induced by the filtered force $\delta u(r)$ depends on the radial distance from the center of the disk. Note that the shed circulation and induced velocity are negative in the coordinate system used.}
\end{center}
\end{figure}

For numerical stability of simulations using actuator models, the body force distribution (along the blades in ALM or across the rotor swept area in ADM) is typically filtered via a convolution kernel~\cite{Calaf2010a, Meyers2010a, Martinez2017b, Martinez2019a}. Unless the geometric footprint of the filtered force properly resolves the smallest scales of the shed vorticity, the induced velocity defect is under-predicted, and hence power generation is over-predicted~\cite{Martinez2017b, Martinez2019a}. A very finely resolved actuator disk will shed a vortex cylinder of infinitesimal thickness at the edge of the disk~\cite{Burton2011a}, as shown in Figure~\ref{fig:diagram}a. At the coarser resolutions typically used in simulations, an ADM under-predicts the velocity gradient at the edge of the disk when compared to experiments~\cite{Lignarolo2016a}, indicating under-resolved vorticity generation. With only a few points across the rotor, the under-prediction of the induced velocity distribution results in an over-prediction of power by 10\% or more~\cite{Munters2017a}. An alternative ADM formulation that uses the Euler equations to directly model the infinitesimal vortex rings emanating from the actuator disk, mitigates these issues and has power coefficient errors of less than 1\%~\cite{vanKuik2016a}. However, such simulations cannot model the turbulent breakdown of the wake or be embedded in large eddy simulations (LES). Clearly, corrections to ADM are needed to improve power coefficient predictions in LES of wind farms.

In this communication, we develop a correction factor for the power production of filtered actuator disks by examining the structure of the shed vorticity. First, we derive a vortex cylinder model for filtered actuator disks that predicts the disk-averaged velocity generated by the shed vorticity and compare these theoretical predictions to simulations. Then, we derive a correction factor for use in simulations that collapses the power coefficient measured at a range of grid resolutions and body force filter widths onto the power curve predicted by axial momentum theory. Recent work on using similar considerations to derive corrections for the ALM approach have led to the filtered actuator line model and improved results for ALM-LES~\cite{Martinez2017b, Martinez2019a}.

\section{The filtered actuator disk}
\label{sec:filtered-adm}
We follow the filtered local actuator disk formulation of Meyers \& Meneveau\cite{Meyers2010a} and Calaf \textit{et al.}\cite{Calaf2010a}, which writes the total thrust force $F$ in terms of the local thrust coefficient $C_T'$, air density $\rho$, radius of the rotor swept area $R$, and disk averaged velocity $u_d$ as
\begin{equation}
F = -\frac{1}{2} \rho \pi R^2 C_T' u_d^2.
\end{equation}
This thrust force is directed in the unit normal direction of the disk $\mathbf{\hat{x}}$ and is spatially distributed with a normalized (integrates to one) indicator function $\mathcal{R}(\mathbf{x})$
\begin{equation}
\mathbf{f}(\mathbf{x}) = F \mathcal{R}(\mathbf{x}) \,\mathbf{\hat{x}}.
\end{equation}
The disk averaged velocity is expressed as a weighted average of the normal component of the velocity field $\mathbf{u}(\mathbf{x})$
\begin{equation}
\label{eq:disk-average}
u_d = \int \mathcal{R}(\mathbf{x}) \, \mathbf{u}(\mathbf{x}) \cdot \mathbf{\hat{x}} \, d^3\mathbf{x},
\end{equation}
with $\mathcal{R}(\mathbf{x})$ used as the weighting function.

The normalized indicator function for a disk of finite thickness $s$ can be written with units of inverse volume as
\begin{equation}
\label{eq:sharp-indicator}
\mathcal{I}(\mathbf{x}) = \frac{1}{s \pi R^2}\left[ H(x+s/2) - H(x - s/2) \right] H(R - r),
\end{equation}
where  $r$ is the radial distance along the disk written in terms of the transverse coordinate directions $r^2 = y^2 + z^2$ and $H(x)$ is the Heaviside function. In order to avoid numerical problems, such as unphysical oscillations and instabilities, that arise when applying sharp forces on grid points, a smoothed indicator function, obtained by convolving $\mathcal{I}(\mathbf{x})$ with a filtering kernel $G(\mathbf{x})$
\begin{equation}
\label{eq:smooth-indicator}
\mathcal{R}(\mathbf{x}) = \int G(\mathbf{x}-\mathbf{x'}) \, \mathcal{I}(\mathbf{x'}) \, d^3\mathbf{x'}
\end{equation}
is typically used. In most implementations~\cite{Calaf2010a, Meyers2010a}
\begin{equation}
G(\mathbf{x}) = \left(\frac{6}{\pi \Delta^2}\right)^{3/2} \exp \left( -\frac{6\lVert\mathbf{x}\rVert^2}{\Delta^2} \right)
\end{equation}
is a Gaussian filtering kernel whose second moment is the same as a box filter of size $\Delta$~\cite{Pope2000a}. The filter width $\Delta= \alpha h $ is the product of a factor of order unity $\alpha$ and the effective grid size\footnote{The effective grid size is expressed as the geometric mean $h = (\Delta x \Delta y \Delta z)^{1/3}$ of the grid spacings in some formulations~\cite{Lignarolo2016a}.} $h = \sqrt{\Delta x^2 + \Delta y^2 + \Delta z^2}$, where $\Delta x$, $\Delta y$, and $\Delta z$ are the grid spacings.

For the Gaussian-filtered indicator function $\mathcal{R}(\mathbf{x})$, there exists a convenient decomposition that splits the indicator function into a product of normal and radial components
\begin{equation}
\mathcal{R}(\mathbf{x}) = \mathcal{R}_1(x) \mathcal{R}_2(r).
\end{equation}
The normal component 
\begin{equation}
\mathcal{R}_1(x) = \frac{1}{s} \left(\frac{6}{\pi \Delta^2}\right)^{1/2} \int \left[ H\left(x'+\frac{s}{2}\right) - H\left(x' - \frac{s}{2}\right) \right] \exp \left( -6\frac{(x-x')^2}{\Delta^2} \right) \, d x' 
\end{equation} 
has the analytic solution
\begin{equation}
\mathcal{R}_1(x) = \frac{1}{2s}\left[\mathrm{erf}\left(\frac{\sqrt{6}}{\Delta}\left(x+\frac{s}{2}\right) \right) - \mathrm{erf}\left(\frac{\sqrt{6}}{\Delta}\left(x-\frac{s}{2}\right) \right)\right].
\end{equation}
The radial component
 \begin{equation} 
 \label{eq:smoothed-indicator-2}
 \mathcal{R}_2(r) = \frac{1}{\pi R^2}\frac{6}{\pi \Delta^2} \int \! \! \! \int H\left(R - \sqrt{y'^2+z'^2}\right) \exp \left( -6\frac{(y-y')^2 + (z-z')^2}{\Delta^2} \right) \, d y' \, d z'
\end{equation}
can be computed numerically or evaluated analytically when $\Delta/R <<1$ (see Section~\ref{sec:correction}). This decomposition proves useful for implementing the smoothed indicator function~\eqref{eq:smooth-indicator} in simulations and computing the vortex cylinder model discussed in the following section.

\section{A vortex cylinder theory for filtered actuator disks}
\label{sec:filtered-adm-theory}
To develop the vortex cylinder theory for filtered actuator disks, we use results from blade element methods and the vortex cylinder model~\cite{Burton2011a} with modifications to include the indicator function from the filtered actuator disk. For each radial element of the disk, the lift on the blade per unit radius is given by
\begin{equation}
l(r) = \frac{1}{2}\rho \pi R^2 C_T' u_d^2 \mathcal{R}_2(r) 2 \pi r.
\end{equation}
For a large rotational speed $\Omega$, the relative velocity over the blade is $V = \Omega r$. 
Applying the Kuta-Joukowsky theorem~\cite{Milne-Thomson1973a} with this relative velocity, the circulation distribution is
\begin{equation}
\Gamma(r) = \frac{l(r)}{\rho V} = \frac{\pi^2 R^2 C_T' u_d^2 \mathcal{R}_2(r)}{\Omega}.
\end{equation}

The annular lift and circulation distribution will shed concentric semi-infinite vortex cylinders, as shown in Figure~\ref{fig:diagram}b. The circulation strength per unit radius $d \Gamma/dr$ is shed once per rotation of the rotor. During one revolution of the rotor, the wake will travel a distance $\ell_\text{adv}$. Assuming that the wake moves uniformly at the disk velocity $u_d$, this advection distance is
\begin{equation}
\ell_\text{adv} = u_d \frac{2\pi}{\Omega}.
\end{equation}
Therefore, the shed circulation strength per unit area is
\begin{equation}
\label{eq:circulation-distribution}
\eta(r) = \frac{d\Gamma}{dr} \frac{1}{\ell_\text{adv}} = \frac{\pi R^2 C_T' u_d }{2} \frac{d \mathcal{R}_2}{dr}.
\end{equation}

For a single semi-infinite vortex cylinder with infinitesimal thickness and circulation strength per unit area of $\eta(r) = \eta_0 \delta(r-R)$, shown in Figure~\ref{fig:diagram}a, the only non-vanishing component of the axial induced velocity is within the cylinder itself~\cite{Gibson1974a}. At the start of the disk, the axial induced velocity within the cylinder is uniformly $\eta_0/2$~\cite{Gibson1974a}. For the circulation distribution in~\eqref{eq:circulation-distribution}, we can superimpose the induced velocities for each of the shed concentric semi-infinite vortex cylinders. At radial location $r$, the only contribution to the induced velocity comes from cylinders with a radius $r'$ greater than $r$. Integrating the contributions and noting that $\lim_{r \rightarrow \infty}\mathcal{R}_2(r) = 0$, the induced velocity $\delta u(r)$ at the rotor plane and along the radius is given by
\begin{equation}
\label{eq:induced-velocity}
\frac{\delta u(r)}{U_\infty} = \frac{1}{2U_\infty}\int_r^\infty \eta(r) \, dr = - \pi R^2 \frac{C_T'}{4} \frac{u_d}{U_\infty} \mathcal{R}_2(r).
\end{equation}
As in ~\eqref{eq:disk-average}, the disk-averaged velocity can now be found by integrating the velocity at the disk $U_\infty + \delta u(r)$ weighted by the radial indicator function
\begin{align}
\label{eq:filtered-ud-start}
\frac{u_d}{U_\infty} &= \int_0^\infty \left[1+\delta u(r)\right] \mathcal{R}_2(r) 2 \pi r \, dr = 1 - \frac{u_d}{U_\infty} \left(\pi R^2 \frac{C_T'}{4} \int_0^\infty \mathcal{R}^2_2(r) 2 \pi r \, dr \right) \\
\label{eq:filtered-ud}
\frac{u_d}{U_\infty}  &= \left(1 +  \pi R^2 \frac{C_T'}{4}\int_0^\infty \mathcal{R}^2_2(r) 2 \pi r \, dr\right)^{-1}.
\end{align}

In the limit of a vanishing filter width $\Delta \rightarrow 0$, the disk-averaged velocity of a filtered actuator disk~\eqref{eq:filtered-ud} tends to the solution of standard axial momentum theory, where the disk averaged velocity $u_d / U_\infty = (1-a)$~\cite{Burton2011a} and local thrust coefficient $C_T' = 4a(1-a)$~\cite{Calaf2010a, Meyers2010a} are a function of the induction factor $a$. In this limit, the filtered indicator function tends to a Heaviside function, i.e. $\lim_{\Delta \rightarrow 0} \mathcal{R}_2(r) = \pi^{-1}R^{-2}H(R-r)$. Therefore, the disk-averaged velocity goes to
\begin{equation}
\lim_{\Delta \rightarrow 0} \frac{u_d}{U_\infty} =\left( 1 + \frac{C_T'}{4}\right)^{-1} = \left( 1 + \frac{a}{1-a}\right)^{-1} = (1-a),
\end{equation}
as predicted by axial momentum theory~\cite{Burton2011a}. 

The disk-averaged velocity in~\eqref{eq:filtered-ud} is compared to simulations of actuator disks under uniform inflow using the pseudo-spectral code LESGO~\cite{Stevens2018a}. The inflow velocity $U_\infty$ is prescribed using a fringe region forcing that spans 25\% of the the domain~\cite{Stevens2014a}. The actuator disk with diameter $D=2R$ is placed in a domain $L_x = 15.36D$ long with a cross-section with dimensions $L_y = L_z = 5.76D$. The center of the disk is located $3D$ from the inlet of the domain and in the center of the cross-section. All simulations have the same grid cell aspect ratio of $\Delta x=\Delta y = 4\Delta z$. The Smagorinsky subgrid scale model with $C_s = 0.16$ is used for numerical stability; however, the results are not affected by the value of $C_s$. Molecular viscosity is neglected. A number of simulations at various $C_T'$, $\Delta$, and $h$ are run and the power coefficient $C_P = C_T' (u_d/U_\infty)^3$\,\cite{Meyers2010a} is computed from the simulation.

Despite the large domain, the simulations experience some blockage effect with the velocity slightly accelerating at the edges of the domain. To correct for this blockage, we assume that the streamwise pressure gradient will distribute the total mass flow rate lost at the rotor disk $\rho  \pi R^2 (U_\infty - u_d)$ equally across the cross section of the domain; i.e. the velocity far from the rotor is $U_\infty + (U_\infty-u_d) \pi R^2/(L_zL_y)$, very close to what we observe in simulations. To obtain $u_d$ at the rotor, we need to subtract the extra velocity, and so the resulting corrected disk-averaged velocity is $u_d - (U_\infty - u_d) \pi R^2/(L_yL_z)$. This correction is small; even for the largest local thrust coefficient $C_T'=2$, the blockage terms correction on the power coefficient is approximately 3\%, which is much smaller than the largest filter-related power coefficient error.

\begin{figure}[t]
\includegraphics{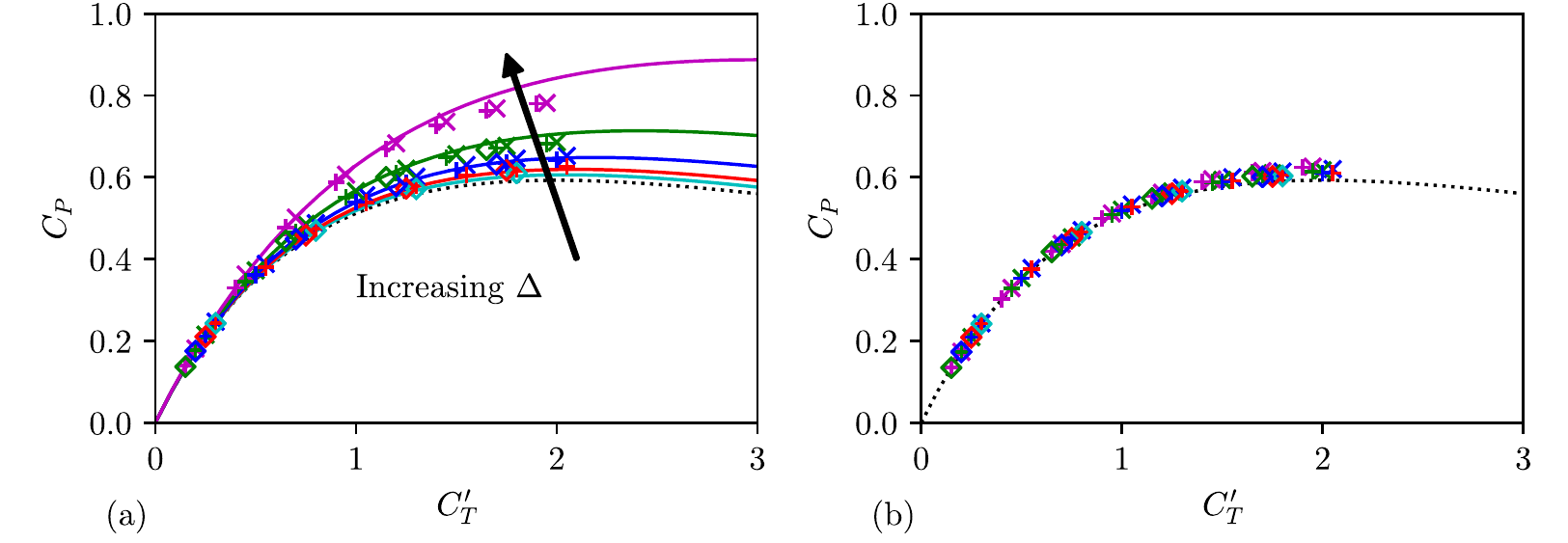}
\caption{(a) Comparison of power coefficient $C_P = C_T' (u_d/U_\infty)^3$ curves for filtered actuator disk simulations (symbols), filtered indicator function theory using~\eqref{eq:filtered-ud} to compute $u_d$ (\full), and axial momentum theory (\dotted). (b) Comparison of the $C_P$ curve for filtered actuator disk simulations with correction factor (symbols) and axial momentum theory (\dotted).  Colors denote filter widths of $\Delta = 0.517D$ (magenta), $\Delta = 0.259D$ (green), $\Delta = 0.129D$ (blue), $\Delta = 0.065D$ (red), and $\Delta = 0.032D$ (cyan). Symbols denote grid sizes of $h = 0.172D$ ($\times$), $h = 0.086D$ ($+$), and $h =  0.043D$ ($\diamond$). }
\label{fig}
\end{figure}

The vortex cylinder theory described herein is compared to axial momentum theory and simulations with various filter widths and grid resolutions in Figure~\ref{fig}a. All simulations over-predict the power coefficient compared to axial momentum theory. The magnitude of this over-prediction is primarily dependent on the filter width $\Delta$ and relatively insensitive to resolution; i.e. for a given filter width $\Delta = \alpha h$, all simulations collapse onto the same curve regardless of $\alpha$. Furthermore, the power curves predicted by the filtered vortex cylinder model largely agree with the simulations, with some discrepancy at coarse resolutions with large local thrust coefficients. 

\section{Correction factor for simulations}
\label{sec:correction}
The theory in Section~\ref{sec:filtered-adm-theory} provides a means for correcting the disk-averaged velocity computed in simulations. Instead of directly averaging using the filtered indicator function, as in~\eqref{eq:disk-average}, we introduce a correction factor $M$ to compute the disk-averaged velocity as
\begin{equation}
\label{eq:disk-average-corrected}
u_d^\text{corr} = M \int \mathcal{R}(\mathbf{x}) \, \mathbf{u}(\mathbf{x}) \cdot \mathbf{\hat{x}} \, d^3\mathbf{x}.
\end{equation}
The correction factor $M$ is found by ensuring that the disk-averaged velocity for the force satisfies the momentum equation
\begin{equation}
\label{eq:ud-adm}
\frac{u_d^\text{corr}}{U_\infty} = (1-a) =  \frac{4}{4+C_T'}.
\end{equation}
Replacing $\int \! \mathcal{R}(\mathbf{x}) \, \mathbf{u}(\mathbf{x}) \cdot \mathbf{\hat{x}} \, d^3\mathbf{x}$ in~\eqref{eq:disk-average-corrected} with~\eqref{eq:filtered-ud-start} expressed using the corrected disk-averaged velocity $u_d^\text{corr}$
\begin{equation}
\frac{u_d^\text{corr}}{U_\infty} = M \left(1 -  \pi R^2 \frac{C_T'}{4} \frac{u_d^\text{corr}}{U_\infty}\int_0^\infty \mathcal{R}^2_2(r) 2 \pi r \, dr\right),
\end{equation}
and replacing $u_d^\text{corr}/U_\infty$ with~\eqref{eq:ud-adm}, we can solve for the correction factor 
\begin{equation}
\label{eq:M}
M^{-1} = 1 + \frac{C_T'}{4} \left( 1- \pi R^2 \int_0^\infty \mathcal{R}^2_2(r) 2 \pi r \, dr \right).
\end{equation}

Figure~\ref{fig}b compares $C_P$ computed using $u_d$ from simulations evaluated as in~\eqref{eq:disk-average-corrected} and with M given by~\eqref{eq:M} to the result of axial momentum theory. Except for the correction factor, the simulations are identical to those of Section~\ref{sec:filtered-adm-theory}. As already noted in Section~\ref{sec:filtered-adm-theory}, for the uncorrected simulations, the power coefficient depends on the filter width and in cases with large filter widths, greatly exceeds the Betz limit. After applying the correction factor in ~\eqref{eq:disk-average-corrected}, the dependence on the filter width is eliminated. Furthermore, the power measurements collapse onto a single curve that differs only slightly from the axial momentum theory predictions.  

\begin{figure}[t]
\begin{center}
\includegraphics{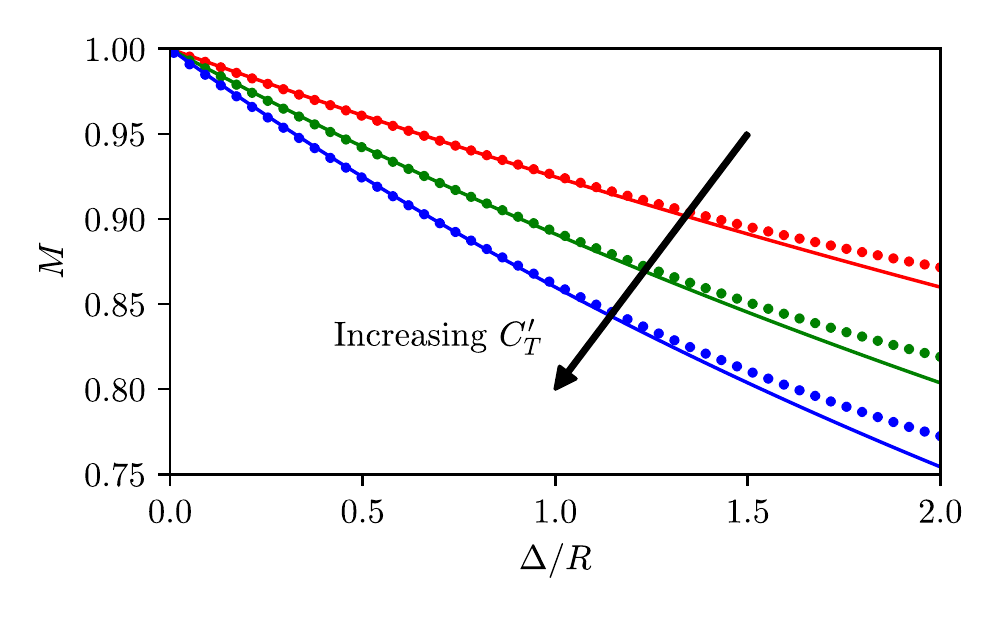}
\end{center}
\caption{Correction factors~\eqref{eq:M} computed numerically ($\cdot$) compared to the approximate expression~\eqref{eq:M-approx} (\full) for $C_T'=1$ (red), $C_T'=3/2$ (green), and $C_T' = 2$ (blue). }
\label{fig:integral}
\end{figure}

For small filter sizes, we can expand the integral in~\eqref{eq:M} using a Taylor series expansion in $\Delta$ around $\Delta = 0$ and recalling that the integral is unity at $\Delta = 0$
\begin{equation}
\label{eq:taylor-series}
\pi R^2 \int_0^\infty \mathcal{R}^2_2(r) 2 \pi r \, dr = 1 + \Delta \frac{d}{d\Delta} \left[\pi R^2 \int_0^\infty \mathcal{R}^2_2(r) 2 \pi r \, dr \right]_{\Delta = 0} + \mathcal{O}(\Delta^2).
\end{equation}
When $\Delta/R <<1$, the curvature of the disk is negligible when evaluating the convolution~\eqref{eq:smoothed-indicator-2}, and the solution for a Gaussian filter is equal to the one-dimensional convolution of the Gaussian with the Heaviside function, i.e. $\mathcal{R}_2(r) = \frac{1}{\pi R^2} \frac{1}{2} \left[ 1 - \mathrm{erf}\left(\sqrt{6}\frac{r-R}{\Delta}\right)\right]$. Replacing this equation into~\eqref{eq:taylor-series}, making a change of variables $\xi = \sqrt{6} (r-R)/\Delta$
\begin{equation}
\frac{d}{d\Delta} \pi R^2 \int_0^\infty \mathcal{R}^2_2(r) 2 \pi r \, dr  = \sqrt{\frac{2}{3\pi}}\frac{1}{R} \int_{-\frac{R\sqrt{6}}{\Delta}}^\infty \left[ 1 - \mathrm{erf}(\xi)\right] \xi e^{-\xi^2} \left( 1 +\frac{\Delta}{\sqrt{6}R} \xi\right) \, d \xi,
\end{equation}
and evaluating at $\Delta = 0$ yields $\sqrt{\frac{2}{3\pi}}\frac{1}{R} \int_{-\infty}^\infty \left[ 1 - \mathrm{erf}(\xi)\right] \xi e^{-\xi^2} \, d \xi = - \frac{1}{R\sqrt{3\pi}}$. Therefore, for small $\Delta/R$ the correction factor can be written as
\begin{equation}
\label{eq:M-approx}
M = \left(1 + \frac{C_T'}{4}\frac{1}{\sqrt{3\pi}}\frac{\Delta}{R}\right)^{-1}.
\end{equation}
The approximate correction factor is compared to numerical integration of~\eqref{eq:M} in Figure~\ref{fig:integral}, showing that the approximation of $M$ using the Taylor series expansion of the integral in~\eqref{eq:M} is valid up to $\Delta/R = 1.25$ and widely applicable for most LES grids.

\section{Conclusions}
The filtered ADM, which is often used in simulations of large wind farms, over-predicts power production, particularly at coarse grid resolutions. This discrepancy arises because the filtered body force does not adequately resolve the semi-infinite vortex cylinder shed behind the actuator disk, which in turn under-estimates the velocity defect induced at the rotor plane. The theory derived in this communication accurately predicts the disk-averaged velocity induced by the concentric semi-infinite vortex cylinders shed behind the turbine when compared to numerical simulations. The theory then provides a means for correcting the disk-averaged velocity in LES, which return the expected power curve from axial-momentum theory for various grid resolutions and filter widths.

\section*{Acknowledgements}
The authors acknowledge funding from the National Science Foundation (grant CMMI 1635430). Computations made possible by the Maryland Advanced Research Computing Center (MARCC). 

\bibliography{}

\end{document}

%% file: diagram.pdf_tex
\begingroup%
  \makeatletter%
  \providecommand\color[2][]{%
    \errmessage{(Inkscape) Color is used for the text in Inkscape, but the package 'color.sty' is not loaded}%
    \renewcommand\color[2][]{}%
  }%
  \providecommand\transparent[1]{%
    \errmessage{(Inkscape) Transparency is used (non-zero) for the text in Inkscape, but the package 'transparent.sty' is not loaded}%
    \renewcommand\transparent[1]{}%
  }%
  \providecommand\rotatebox[2]{#2}%
  \ifx\svgwidth\undefined%
    \setlength{\unitlength}{468bp}%
    \ifx\svgscale\undefined%
      \relax%
    \else%
      \setlength{\unitlength}{\unitlength * \real{\svgscale}}%
    \fi%
  \else%
    \setlength{\unitlength}{\svgwidth}%
  \fi%
  \global\let\svgwidth\undefined%
  \global\let\svgscale\undefined%
  \makeatother%
  \begin{picture}(1,0.34615385)%
    \put(0.51088916,0.14440847){\color[rgb]{0,0,0}\makebox(0,0)[lb]{\smash{$U_\infty$}}}%
    \put(0.14941172,0.23271869){\color[rgb]{0,0,0}\makebox(0,0)[lb]{\smash{${\color{blue}-\delta u}$}}}%
    \put(0,0){\includegraphics[width=\unitlength,page=1]{diagram.pdf}}%
    \put(0.51988344,0.08748422){\color[rgb]{0,0,0}\makebox(0,0)[lb]{\smash{${\color{blue}-\delta u(r)}$}}}%
    \put(-0.00044637,0.01423372){\color[rgb]{0,0,0}\makebox(0,0)[lb]{\smash{(a) Actuator disk}}}%
    \put(0,0){\includegraphics[width=\unitlength,page=2]{diagram.pdf}}%
    \put(0.36778845,0.24479245){\color[rgb]{0,0,0}\makebox(0,0)[lb]{\smash{$x$}}}%
    \put(0.27604164,0.27924662){\color[rgb]{0,0,0}\makebox(0,0)[lb]{\smash{${\color{red}-\eta_0}$}}}%
    \put(0.14329997,0.09569458){\color[rgb]{0,0,0}\makebox(0,0)[lb]{\smash{$r$}}}%
    \put(0,0){\includegraphics[width=\unitlength,page=3]{diagram.pdf}}%
    \put(0.49949119,0.01423372){\color[rgb]{0,0,0}\makebox(0,0)[lb]{\smash{(b) Filtered actuator disk}}}%
    \put(0.86772602,0.24479245){\color[rgb]{0,0,0}\makebox(0,0)[lb]{\smash{$x$}}}%
    \put(0.81444022,0.27924662){\color[rgb]{0,0,0}\makebox(0,0)[lb]{\smash{${\color{red}-\eta(r)}$}}}%
    \put(0.64966434,0.10211713){\color[rgb]{0,0,0}\makebox(0,0)[lb]{\smash{$r$}}}%
    \put(0,0){\includegraphics[width=\unitlength,page=4]{diagram.pdf}}%
    \put(0.01088797,0.14440859){\color[rgb]{0,0,0}\makebox(0,0)[lb]{\smash{$U_\infty$}}}%
    \put(0,0){\includegraphics[width=\unitlength,page=5]{diagram.pdf}}%
  \end{picture}%
\endgroup%